\documentclass[12pt]{article}
\usepackage[utf8]{inputenc}
\usepackage[english]{babel}
\usepackage{amsmath}
\usepackage{mathrsfs}
\usepackage{amssymb}
\usepackage{amsthm}
\usepackage{bm}
\usepackage{amsfonts}
\usepackage{xspace}
\usepackage[normalem]{ulem}
\usepackage{graphicx}
\usepackage{multirow}
\usepackage{booktabs}
\usepackage{appendix}
\usepackage{enumerate}
\usepackage{url} % not crucial - just used below for the URL
\usepackage{authblk}
\usepackage{xcolor}
\usepackage{subcaption}
\usepackage{xr}
\usepackage{adjustbox}
\usepackage[colorlinks=true,linkcolor=red,citecolor=blue]{hyperref}
\usepackage{enumitem}
 \usepackage{natbib}

\newcommand{\blind}{1}

\makeatletter
\newcommand*{\addFileDependency}[1]{% argument=file name and extension
\typeout{(#1)}% latexmk will find this if $recorder=0
\@addtofilelist{#1}
\IfFileExists{#1}{}{\typeout{No file #1.}}
}\makeatother
\newcommand{\balpha}{\boldsymbol{\alpha}}
\newcommand{\bbeta}{\boldsymbol{\beta}}

\newcommand{\bgamma}{\boldsymbol{\gamma}}
\newcommand{\bX}{\mathbf{X}}

\newcommand{\mP}{\mathbb{P}}

\newcommand{\whp}{\widehat{p}}

\newcommand{\whe}{\widehat{e}}
\newcommand{\whpi}{\widehat{\pi}}
\newcommand{\whm}{\widehat{m}}

\newtheorem{assumption}{Assumption}
\newtheorem{theorem}{Theorem}

\theoremstyle{remark}
\newtheorem{remark}{Remark}

\begin{document}

\def\spacingset#1{\renewcommand{\baselinestretch}%
{#1}\small\normalsize} \spacingset{1}

%%%%%%%%%%%%%%%%%%%%%%%%%%%%%%%%%%%%%%%%%%%%%%%%%%%%%%%%%%%%%%%%%%%%%%%%%%%%%%

\if1\blind
{
  \title{\bf Semiparametric principal stratification analysis beyond monotonicity}
\author{\textbf{Jiaqi Tong$^{1,4}$, Brennan Kahan$^2$, Michael O. Harhay$^{2,3}$ and Fan Li$^{1,4,*}$}

$^1$Department of Biostatistics, Yale University of Public Health, New Haven, Connecticut, USA

$^2$MRC Clinical Trials Unit at UCL, Institute of Clinical Trials and Methodology, London, UK

$^{3}$Department of Biostatistics, Epidemiology and Informatics, Perelman School of Medicine, 
University of Pennsylvania, Philadelphia, PA, USA

$^{4}$Center for Methods in Implementation and Prevention Science, Yale School of Public Health, New Haven, CT, USA

%$^{5}$Clinical and Translational Research Accelerator, Department of Medicine, Yale School of Medicine, New Haven, CT, USA

\emph{*email}: fan.f.li@yale.edu}

  \maketitle
} \fi

\if0\blind
{
  \bigskip
  \bigskip
  \bigskip
  \begin{center}
    {\LARGE\bf Semiparametric principal stratification analysis beyond monotonicity}
\end{center}
  \medskip
} \fi

\bigskip
\begin{abstract}
Intercurrent events, common in clinical trials and observational studies, affect the existence or interpretation of final outcomes. Principal stratification addresses this challenge by defining local average treatment effect estimands within subpopulations, but often relies on restrictive assumptions such as monotonicity and counterfactual intermediate independence. To overcome these limitations, we propose a semiparametric framework for principal stratification analysis leveraging a margin-free, conditional odds ratio sensitivity parameter. Under principal ignorability, we derive nonparametric identification formulas and efficient estimation methods, including a conditionally doubly robust parametric estimator and a debiased machine learning estimator with data-adaptive nuisance learners. Our simulations show that incorrectly assuming monotonicity can frequently lead to biased inference, but incorrectly assuming non-monotonicity when monotonicity holds may maintain approximately valid inference. We demonstrate our methods in the context of a critical care trial, where monotonicity is unlikely to be valid.
\end{abstract}

\noindent%
{\it Keywords:}  Causal inference; efficient influence function; conditional double robustness; odds ratio; sensitivity analysis; intercurrent events.
\vfill

\newpage
\spacingset{1.45} % DON'T change the spacing!

\section{Introduction}
\label{s:intro}
Intercurrent events are post-treatment events (also referred to as intermediate outcomes) that affect either the interpretation or existence of the final outcome, posing a critical challenge to the analysis of clinical trials and observational studies \citep{KahanICH}. A typical example is truncation by death, where mortality occurring post-randomization can lead to ambiguously defined final outcomes \citep{rubin2006causal}. Direct regression adjustment for an observed intercurrent event often fails to produce causally interpretable treatment effects due to post-randomization selection bias. With an intermediate outcome, the principal stratification approach frames the counterfactual intermediate outcomes as a pre-treatment covariate \citep{frangakis2002principal}, and defines the principal causal effects to quantify the average treatment effects among subpopulations. The ICH E9 (R1) addendum has also recently recognized principal stratification as a viable strategy for addressing intercurrent events in clinical trials \citep{ema2020}. 

%% Literature review 

Existing methods for principal stratification analysis typically rely on the monotonicity assumption for nonparametric identification \citep[e.g.,][]{shepherd2006sensitivity,DingandLu2016}. In the context of truncation by death, monotonicity implies that individuals who survive under the control condition must survive under the treatment condition and rules out the ``{harmed}'' stratum. This assumption, however, is controversial in the presence of two active treatments. An alternative assumption---the counterfactual intermediate independence---requires the independence between the pair of counterfactual intermediate outcomes given measured baseline covariates \citep{hayden2005estimator}. This assumption does not restrict the set of principal strata, yet its plausibility has been questioned because capturing all relevant baseline predictors of a post-baseline variable is generally unattainable \citep{2024criticizeVanstee}.

Several previous studies have sought to relax these stringent assumptions. \citet{roy2008principal} introduced a modified monotonicity assumption with two active treatments, and proposed a model-based sensitivity approach in the context of noncompliance. Fully relaxing monotonicity, \citet{DingandLu2016,JiangJRSSB2022,cheng2023multiply} and \citet{tong2024doubly} developed bias-corrected principal causal effect estimators with sensitivity functions parameterized by the ratio of strata proportions. {\cite{wu2024quantifying} proposed a sensitivity analysis approach based on the Pearson correlation coefficient and derived tighter bounds for the  principal score.} Yet, these existing sensitivity functions have implicitly encoded partial knowledge of the true marginal distribution of the counterfactual intermediate outcomes. The range of the sensitivity function is thus not margin-free and constrained by the unknown marginal distribution of the intermediate outcome (c.f. Proposition 5 in \cite{DingandLu2016} {and Proposition 3 in \cite{wu2024quantifying}}), leading to potential operational challenges. To this end, an unconstrained margin-free sensitivity analysis framework for principal stratification analysis, which unifies the monotonicity assumption and the counterfactual intermediate independence assumption, is of interest. In fact, these two assumptions are mutually exclusive and have been previously presented as competing approaches for principal stratification analysis \citep[e.g.,][]{isenberg2024weighting}. 

%% Contributions of this paper

In this article, we develop a margin-free sensitivity framework for conducting principal stratification analysis with a binary intermediate outcome. Our novel contributions are threefold. First, we introduce an odds ratio sensitivity parameter, possibly covariate-dependent, for the intermediate outcome, offering ease of interpretation and operation. Our odds ratio sensitivity parameter shares a similar flavor with the copula-based sensitivity parameter proposed for handling continuous intermediate outcomes \citep{bartolucci2011modeling, lu2023principal}, yet they diverge (Remark \ref{remark:dis_copula}). Second, we establish the semiparametric efficiency theory that unifies two competing approaches for principal stratification analysis---that assuming monotonicity and that assuming counterfactual intermediate independence. Under the mean principal ignorability assumption, we derive the efficient influence function (EIF) to motivate a conditionally doubly robust estimator for each principal causal effect. Our conditionally doubly robust estimator is parameterized by a set of nuisance functions including the propensity score, principal score, and the outcome mean. Given a correctly specified principal score model, the validity of this estimator necessitates correct specification of either the propensity score model or the outcome mean model, but not necessarily both. Interestingly, when the monotonicity assumption holds and the implied odds ratio parameter converges to infinity, our conditionally doubly robust estimator coincides with the triply robust estimator of \citet{JiangJRSSB2022}. Third, we propose a nonparametric debiased machine learning estimator where each nuisance function is estimated via data-adaptive methods under the cross-fitting scheme \citep{dml}. Under mild regularity conditions, we show that the debiased machine learning estimator achieves $\sqrt{n}$ consistency, asymptotic linearity, and semiparametric efficiency, even when nuisances are estimated at slower rates by data-adaptive methods. Through extensive simulations, we explore the finite-sample performance of the proposed estimators under different data-generating processes and obtain new insights. Finally, we apply the proposed methods to a critical care trial, where monotonicity is questionable due to comparing two active interventions. For practical implementation, we provide the \texttt{PSor} R package with a user guide at \href{https://github.com/deckardt98/PSor}{GitHub}.

\section{Notation and assumptions}
\label{s:model}
We consider a setting with a binary treatment assignment $Z\in\{0,1\}$, a binary intermediate outcome $D\in\{0,1\}$ and a final outcome $Y\in\mathbb{R}$. {Although our results are applicable to two active treatments, we refer to $Z=1$ and $Z=0$ as the treatment and control conditions for simplicity.} Let $n$ be the sample size, and for each unit, we observe a data vector $\mathcal{O}\equiv\{Y,D,Z,\bX\}$, where $\bX$ is a vector of pre-treatment covariates. We define $Y(z)$ and $D(z)$ as the potential final outcome and potential intermediate outcome that would have been observed under assignment $z$, respectively. Under the Stable Unit Treatment Value Assumption, we have $Y=ZY(1)+(1-Z)Y(0)$ and $D=ZD(1)+(1-Z)D(0)$. Principal stratification analysis assumes that the joint potential intermediate outcomes---$\{D(0),D(1)\}$---can be treated as a pre-treatment covariate \citep{frangakis2002principal}. We define the principal strata as $\{D(0),D(1)\}=d_0d_1$ if $D(0)=d_0$ and $D(1)=d_1$ with $d_0,d_1\in\{0,1\}$. For example, under noncompliance, $\{01,11,10,00\}$ correspond to compliers, always-takers, defiers, and never-takers \citep{angrist1994}; under truncation by death, these strata correspond to protected, always-survivors, harmed, and never-survivors \citep{rubin2006causal}. The principal causal effects are defined as $\mu_{d_0d_1}=E\{Y(1) - Y(0) | D(0) = d_0, D(1) = d_1\}$. As a typical example, the complier average causal effect (CACE) within stratum $01$ is of interest for quantifying treatment efficacy; the survivor average causal effect (SACE) within stratum $11$ is of primary interest when $Y$ is a final non-mortality outcome that may be truncated by the intermediate survival status. However, since $\{D(0), D(1)\}$ is not fully observable, additional assumptions are needed to point identify $\mu_{d_0d_1}$.
\begin{assumption}[\emph{Treatment ignorability}]\label{assump:treatment_ignorability}
 %The treatment assignment is unconfounded such that 
 $Z\perp \{D(0),D(1),Y(0),Y(1)\}|\bX$; furthermore, the propensity score $\pi(\bX)=\Pr(Z=1|\bX)$ is uniformly bounded.
\end{assumption}
\begin{assumption}[\emph{Conditional odds ratio}]\label{assump:conditional_Ind_potential D}
Defining $\mathcal{X}$ as the covariate support and $e_{d_0d_1}(\bX)\equiv\Pr(D(0)=d_0,D(1)=d_1|\bX)$ as the principal score, there exists a known function $\theta(\bullet):\mathcal{X}\to[0,\infty]$, such that $\{e_{11}(\bX)e_{00}(\bX)\}/\{e_{10}(\bX)e_{01}(\bX)\}=\theta(\bX)$.
\end{assumption}
Assumption \ref{assump:treatment_ignorability} is standard to rule out unmeasured confounding, and holds marginally without $\bX$ in randomized trials. {Assumption \ref{assump:conditional_Ind_potential D} introduces an odds ratio sensitivity parameter $\theta(\bX)$ to quantify the conditional correlation between the counterfactual intermediate outcomes $\{D(0),D(1)\}$ given baseline covariates. Though $\theta(\bX)$ is a cross-world measure and not directly estimable, its magnitude can be interpreted as follows (using truncation by death as an example). When $\theta(\bX) > 1$, the odds of $D(1)=1$ given $D(0)=1$ are higher than those given $D(0)=0$, suggesting that patients are more likely to survive under the treated condition if they would also have survived under the control condition. Conversely, when $\theta(\bX) < 1$, the treatment is more likely to reduce survival among patients who would have survived under the control condition. Neither case assumes strict monotonicity.}

We notice that using odds ratio as a sensitivity parameter to relax monotonicity has been previously discussed by \citet{jemiai2005semiparametric} and \citet{shepherd2008does,shepherd2011sensitivity}. While they pursued a constant log odds ratio parameter to study principal causal effects, Assumption \ref{assump:conditional_Ind_potential D} is more general with a conditional odds ratio that may depend on covariates. In general, the motivation for pursuing Assumption \ref{assump:conditional_Ind_potential D} is twofold. First, the odds ratio parameter is widely reported in epidemiology and biomedical research, making it at least a familiar quantity for practitioners. Second, different from existing sensitivity methods in \citet{DingandLu2016} and \citet{JiangJRSSB2022}, the odds ratio parametrization is margin-free such that the range of $\theta(\bX)$ does not depend on the true margins, as long as the margins are strictly positive. We offer two additional remarks below. 

\begin{remark}[\emph{Distinction from copula}]\label{remark:dis_copula}
With a continuous intermediate outcome, a copula approach has been introduced in the principal stratification literature for margin-free sensitivity analysis \citep{bartolucci2011modeling,jiang2021identification,lu2023principal}. However, this approach is not directly applicable to discrete data \citep{Genest_Nešlehová_2007}. The reason is that Sklar’s theorem, a cornerstone of the copula framework, does not ensure the uniqueness of the copula function for discrete margins. Instead, only a subcopula---identifiable at the jumps of the joint distribution---is determined \citep{nelsen2006introduction}. Without relying on Sklar’s theorem, the odds ratio has been shown to be the unique (up to one-to-one mapping) margin-free parametrization for binary data (Theorem 6.3 in \cite{rudas_lectures_2018}), making it a natural sensitivity parameter to characterize the association between the binary potential intermediate outcomes $D(0)$ and $D(1)$.
\end{remark}

\begin{remark}[\emph{Alternative association measures}]
{Supplementary Material Table S2 summarizes alternative association measures for $D(0)$ and $D(1)$---including the probability ratio (used in \cite{DingandLu2016} and \citet{JiangJRSSB2022}), the risk ratio, and the Pearson correlation coefficient (i.e., the $\phi$ coefficient used in \citet{wu2024quantifying})---that can also be used for sensitivity analysis, but they are not margin-free. In contrast, based on Theorem 6.3 in \cite{rudas_lectures_2018}, the Yule’s coefficient of colligation \citep{YulesCoefficient} has a one-to-one mapping to the odds ratio and is therefore margin-free. Consequently, Assumption \ref{assump:conditional_Ind_potential D} and our subsequent results can be equivalently stated using the coefficient of colligation.}
\end{remark}

{Assumption \ref{assump:conditional_Ind_potential D} includes monotonicity and counterfactual intermediate independence as special cases. First, $\theta(\bX)=1$ is equivalent to $D(0)\perp D(1)|\bX$, in which case the counterfactual intermediate independence in \citet{hayden2005estimator} holds. We next address the more nuanced case of monotonicity. It is evident that $\theta(\bX)\to\infty$ is implied by either $e_{10}(\bX) = 0$ or $e_{01}(\bX) = 0$ for all $\bX\in\mathcal{X}$. However, $\theta(\bX)\to\infty$ alone is not sufficient to fully characterize monotonicity. To see this, suppose that we have a binary covariate with $X\in\{0,1\}$, and Table \ref{tab:combined_contingency} specifies the conditional distribution of $\{D(0),D(1)\}$ given $X$. Although both $\theta(0)$ and $\theta(1)$ equal $\infty$, the dominance relationship now depends on the value of $X$ because $D(1)\geq D(0)$ under $X=0$ and $D(1)\leq D(0)$ under $X=1$. More generally, $\theta(\bX) \to \infty$ occurs if either compliers or defiers are ruled out for certain levels of $\bX$; and therefore an additional restriction is needed to precisely characterize monotonicity.}

\begin{table}[ht]
\centering
\caption{An example conditional distribution of $\{D(0),D(1)\}$ given a binary covariate.}
\label{tab:combined_contingency}
\begin{tabular}{|c|c|cc|c|c|}
\hline
& & $D(0)=0$ & $D(0)=1$ & $\theta(X)$ & $D(1)\geq D(0)?$\\
\hline
\multirow{2}{*}{$X=0$} & $D(1)=0$ & $e_{00}(X=0)=1/3$ & $e_{10}(X=0)=0$ & \multirow{2}{*}{$\infty$}&\multirow{2}{*}{$\checkmark$} \\
& $D(1)=1$ & $e_{01}(X=0)=1/3$ & $e_{11}(X=0)=1/3$ & &\\
\hline
\multirow{2}{*}{$X=1$} & $D(1)=0$ & $e_{00}(X=1)=1/3$ & $e_{10}(X=1)=1/3$ & \multirow{2}{*}{$\infty$}& \multirow{2}{*}{$\times$}\\
& $D(1)=1$ & $e_{01}(X=1)=0$ & $e_{11}(X=1)=1/3$ & &\\
\hline
\end{tabular}
\end{table}

{To introduce that additional restriction, we let $p_z(\mathbf{X})=E(D|\bX,Z=z)$ denote the conditional expectation of the observed intermediate outcome. Following \citet{JiangJRSSB2022} and with a slight abuse of terminology, we also refer to $p_z(\bX)$ as the principal score, as there is a one-to-one mapping between $\{e_{d_0d_1}(\bX):d_0,d_1\in\{0,1\},\sum_{d_0,d_1}e_{d_0d_1}(\bX)=1\}$ and $\{p_0(\bX),p_1(\bX),\theta(\bX)\}$. Assuming that at most one principal stratum is empty, we show in Section S1 of the Supplementary Material that $\theta(\bX) \to \infty$ and $p_1(\bX) > p_0(\bX)$ for all $\bX\in\mathcal{X}$ are sufficient and necessary conditions for $D(1) \geq D(0)$. To show sufficiency, we note that $p_1(\bX)>p_0(\bX)$ implies $e_{01}(\bX)>e_{10}(\bX)$, and thus for $\theta \to \infty$ to hold, $e_{10}(\bX)=0$ must hold because $e_{01}(\bX)=0$ would imply a contradiction of $e_{10}(\bX)<0$. The necessity  is trivial. To proceed, we let $\theta\in[0,\infty]$ to allow for division by zero; that is if $e_{z,1-z}(\bX)=0$, we simply set $\theta(\bX) \to \infty$. As a final caveat, certain degenerate cases, such as strong monotonicity \citep{DingandLu2016}, warrant more subtle considerations, as elaborated in Section S1 of the Supplementary Material.} 

{In practice when multiple covariates are involved, the true $\theta(\bX)$ may take a complex form requiring a non-trivial number of finer sensitivity parameters, which can complicate both the implementation and interpretation. As a practical approach, one may approximate $\theta(\bX)$ by a constant odds ratio close to its expected value over the distribution of $\bX$. Our limited simulation study in Section \ref{s:simulation} shows that this constant approximation strategy yields only mild bias and undercoverage, while substantially improving upon estimates that incorrectly assume monotonicity. Last but not the least, we state an ignorability assumption required for the point identification.}

\begin{assumption}[\emph{Mean Principal ignorability}]\label{assump:PI_weak}
\begin{align*}
    &\text{3(a): } E\left\{Y(1)|D(0)=1,D(1)=1,\bX\right\}=E\left\{Y(1)|D(0)=0,D(1)=1,\bX\right\}.\\
    &\text{3(b): } E\left\{Y(1)|D(0)=1,D(1)=0,\bX\right\}=E\left\{Y(1)|D(0)=0,D(1)=0,\bX\right\}.\\
    &\text{3(c): } E\left\{Y(0)|D(0)=1,D(1)=1,\bX\right\}=E\left\{Y(0)|D(0)=1,D(1)=0,\bX\right\}.\\
    &\text{3(d): } E\left\{Y(0)|D(0)=0,D(1)=1,\bX\right\}=E\left\{Y(0)|D(0)=0,D(1)=0,\bX\right\}.
\end{align*}
\end{assumption}
Assumption \ref{assump:PI_weak} requires that for $z,z' \in \{0,1\}$, the mean of $Y(1-z)$ is homogeneous across both levels of $D(z)$, conditional on $D(1-z)=z'$ and baseline covariates. In the truncation by death setting, Assumption \ref{assump:PI_weak}(a) indicates that, among those who would have survived under condition $Z=1$, the mean potential outcome under $Z=1$ is conditionally independent of survival status under condition $Z=0$ given the measured covariates. Importantly, with Assumptions \ref{assump:treatment_ignorability} and \ref{assump:conditional_Ind_potential D}, we only need two out of four conditional exchangeability conditions %(referred to as Strong Partial Principal Ignorability, or SPPI) 
to point identify a single PCE; that is, out of the six possible combinations, we need  3(a) and 3(d) for the CACE, 3(a) and 3(c) for the SACE, 3(b) and 3(d) for the PCE among never-takers, and 3(b) and 3(c) for the PCE among defiers. A subset of these combinations has been explored previously. For example, under monotonicity such that $D(1)\geq D(0)$, 3(a) and the combination of 3(a) and 3(c) are known as partial principal ignorability (PPI) and strong PPI, respectively \citep{Zehavi_JRSSA}. %\citet{Zehavi_JRSSA} showed that PPI, along with monotonicity, implies SPPI. 
In addition, 3(d) and the combination of 3(a) and 3(d) are referred to as principal ignorability under one-sided non-compliance and two-sided non-compliance, respectively \citep{feller2017principal}, and the same terminology is used for general principal score methods \citep{DingandLu2016,JiangJRSSB2022}. 
With counterfactual intermediate independence, a combination of 3(a) and 3(d) is referred to as explainable nonrandom noncompliance \citep{Robins1998-ic}, whereas a combination of 3(a) and 3(c) is referred to as explainable nonrandom survival \citep{hayden2005estimator}. For simplicity, we broadly refer to the collection of conditional exchangeability conditions in Assumption \ref{assump:PI_weak} as mean principal ignorability.
%, with an important caveat that point identification of each single PCE requires only two of the four conditions. 

\section{{Estimation based on weighting and regression}}
\label{s:sace}
To simplify the presentation, in what follows, we primarily focus on the truncation by death setting and the SACE estimand, and leave the complete results for the remaining PCEs to Section S4 of the Supplementary Material. Formally, the SACE estimand is defined within the subpopulation of individuals who would survive regardless of treatment assignment, as $\mu_{11}=\mu_{11}^1-\mu_{11}^0$, where for $z\in\{0,1\}$,
\begin{equation}\nonumber%\label{eq:def_saces}
    %\mu_{11}=E\{Y(1)-Y(0)|D(0)=D(1)=1\}=\frac{\displaystyle E\{D(0)D(1)(Y(1)-Y(0))\}}{\displaystyle E\{D(0)D(1)\}},
    \mu_{11}^z=E\{Y(z)|D(0)=D(1)=1\}=\frac{\displaystyle E\{D(0)D(1)Y(z)\}}{\displaystyle E\{D(0)D(1)\}},
\end{equation}
which is identifiable under Assumptions \ref{assump:treatment_ignorability}, \ref{assump:conditional_Ind_potential D} and \ref{assump:PI_weak}(a) and \ref{assump:PI_weak}(c), as explained below.

%\subsection{Weighting and outcome regression estimators}
%\label{sec:saces;ss:weighting}
%In this section, we propose a weighting estimator for SACE, extending the approach originally developed in \cite{hayden2005estimator} for randomized clinical trials. 
%Let $\pi(\bX)\equiv\Pr(Z=1|\bX)$ denote the propensity score. 

Define $m_{zd}(\bX)=E\{Y|Z=z,D=d,\bX\},z,d\in\{0,1\}$ as the mean of the non-mortality outcome conditional on the treatment assignment, the survival status, and covariates. For $\theta(\bX)\neq1$ almost surely, we show in Section S5 of the Supplementary Material that $\mu_{11}$ is point identifiable with
\begin{align}
     \mu_{11}=\frac{E\left\{e_{11}(\bX)ZDY/\left[\pi(\bX)p_1(\bX)\right]\right\}}{E\left\{e_{11}(\bX)ZD/\left[\pi(\bX)p_1(\bX)\right]\right\}}-\frac{E\left\{e_{11}(\bX)(1-Z)DY/\left[(1-\pi(\bX))p_0(\bX)\right]\right\}}{E\left\{e_{11}(\bX)(1-Z)D/\left[(1-\pi(\bX))p_0(\bX)\right]\right\}},\label{eq:saces_identification_weighting}
\end{align}
and, equivalently, with 
\begin{align}
     \mu_{11}=
     \frac{E\{e_{11}(\bX)(m_{11}(\bX)-m_{01}(\bX))\}}{E\{e_{11}(\bX)\}},\label{eq:saces_identification_weighting2}
\end{align}
where 
\begin{align*}
    &e_{11}(\bX)=\frac{1+(\theta(\bX)-1)(p_0(\bX)+p_1(\bX))-\sqrt{\delta(\bX)}}{2(\theta(\bX)-1)},\\
    &\delta(\bX)=\left[1+(\theta(\bX)-1)(p_0(\bX)+p_1(\bX))\right]^2-4\theta(\bX)(\theta(\bX)-1)p_0(\bX)p_1(\bX).
\end{align*}
Here, the principal score $e_{11}(\bX)$ is a function of $\{p_0(\bX), p_1(\bX), \theta(\bX)\}$ through a complex function $\delta(\bX)$. Below, we occasionally write $\mu_{11}(\theta)$ and $e_{11}(\bX;\theta)$ to explicitly acknowledge the dependence on the conditional odds ratio sensitivity function $\theta(\bX)$ when necessary. {To also accommodate the case of monotonicity, we note that $\lim_{\theta \to \infty} e_{11}(\bX;\theta) = (p_0(\bX) + p_1(\bX) - |p_0(\bX) - p_1(\bX)|)/2$. When we place an additional restriction $p_1(\bX) > p_0(\bX)$, this limit reduces to $e_{11}(\bX) = p_0(\bX)$. On the other hand, under counterfactual intermediate independence, we have $e_{11}(\bX) = p_0(\bX)p_1(\bX)$, and that $e_{11}(\bX)$ has a removable discontinuity at a point $\theta=1$ since $\lim_{\theta \to 1} e_{11}(\bX;\theta) = p_0(\bX)p_1(\bX)$. Remark \ref{remark:boudanry} provides additional practical considerations on the two boundary cases.
}

\begin{remark}[\emph{Boundary cases}]\label{remark:boudanry}
{In practice, when addressing counterfactual intermediate independence, we recommend direct specification by setting $e_{11}(\bX) = p_0(\bX)p_1(\bX)$ \citep{hayden2005estimator}. An alternative strategy is to approximate this condition by fixing $\theta(\bX)$ at a value arbitrarily close to one (e.g., $\theta(\bX) = 0.999999$); our simulations indicate that this approximation yields identical results up to 7 decimal places (omitted for brevity). Under the monotonicity assumption, however, we advocate direct specification of principal scores as in \citet{JiangJRSSB2022}, which provides a more principled formulation than attempting to approximate the condition by letting $\theta$ approach $\infty$  (a necessary but not sufficient condition for monotonicity; see Table \ref{tab:combined_contingency}).}
%In practice, to handle the case of counterfactual intermediate independence, we recommend the direct specification by setting $e_{11}(\bX) = p_0(\bX)p_1(\bX)$ as in \citet{hayden2005estimator}. An alternative approach, however, is to use an approximation of $\theta(\bX) = 0.999999$; this approximation has been demonstrated to be highly accurate in our limited simulations. On the other hand, to handle the monotonicity assumption, we recommend the direct specification of principal scores as in \cite{JiangJRSSB2022}. Similarly, an alternative approach is to consider an approximation of setting $\theta$ to be a large number. the limiting form of our own approach as $\theta \to \infty$. Under monotonicity, these two approaches are equivalent.}     
\end{remark}

{To proceed with estimation, we specify parametric models for the propensity score, $\pi(\mathbf{X};\balpha)$, the principal score, $p_z(\bX;\bbeta_z)$, and the outcome mean, $m_{zd}(\bX)=m_{zd}(\bX;\bgamma_{zd})$. We denote the plug-in maximum likelihood estimates of the propensity score, the principal score, and the outcome mean as, $\widehat{\pi}(\bX)=\pi(\bX;\widehat\balpha)$, $\widehat{p}_z(\bX)=p_z(\bX;\widehat\bbeta_z)$, and $\widehat{m}_{zd}(\bX)=m_{zd}(\bX;\widehat\bgamma_{zd})$, respectively. Based on the identification conditions in \eqref{eq:saces_identification_weighting} and \eqref{eq:saces_identification_weighting2}, we propose a weighting estimator ($\widehat{\mu}_{11}^{\text{WT}}$) and an outcome regression estimator ($\widehat{\mu}_{11}^{\text{OR}}$), using empirical averages. For any given $\theta(\bX)$, $\widehat{\mu}_{11}^{\text{WT}}$ and $\widehat{\mu}_{11}^{\text{OR}}$ are consistent if the corresponding nuisance models are correctly specified. To compute interval estimates, we use the robust sandwich variance method described in Section S3 of the Supplementary Material.} To summarize, our approach nests two notable previous contributions as special cases. First, $\widehat{\mu}_{11}^{\text{WT}}$ coincides with the weighting estimator in \cite{hayden2005estimator}, which is derived under counterfactual intermediate independence; Second, $\mu_{11}(\infty)$ in \eqref{eq:saces_identification_weighting} and \eqref{eq:saces_identification_weighting2} under the constraint $p_1(\bX)> p_0(\bX)$ coincides with Theorem 1(a) and (c) in \cite{JiangJRSSB2022}, which are derived under monotonicity.

\section{{Conditionally doubly robust and efficient estimation}}
\label{sec:saces;ss:multiply_robust}
Principal score weighting, although simple, is not optimally efficient. On the other hand, outcome regression requires both the principal score and conditional outcome mean models to be correct for consistent estimation of $\mu_{11}$. To improve efficiency and robustness, we derive the efficient influence function (EIF) under the semiparametric efficiency framework and use the EIF to construct a conditionally doubly robust estimator and an efficient estimator \citep{bickel1993efficient,dml}. 
%In particular, the derivation of the EIF ignores the structural assumptions in a non-parametric sense, which is common in the existing literature \citep{hahn1998role,kennedy2022semiparametric,JiangJRSSB2022}. 
To facilitate the derivation of EIF, we define the following random function of the observed data vector $\mathcal{O}$, for $z\in\{0,1\}$,
\begin{equation*}
    \psi_{F(Y,D,\mathbf{X}),z}=\frac{\mathbf{1}(Z=z)\Big\{F(Y,D,\mathbf{X})-E\{F(Y,D,\mathbf{X})|Z=z,\mathbf{X}\}\Big\}}{\Pr(Z=z|\bX)}+E\{F(Y,D,\mathbf{X})|Z=z,\mathbf{X}\},
\end{equation*}
where $\mathbf{1}(\bullet)$ is the indicator function. Specifically, $\psi_{D,z}$ and $\psi_{YD,z}$ are used to construct the EIF with the following explicit expressions
\begin{align*}
\psi_{D,z}=&\frac{\mathbf{1}(Z=z)}{\Pr(Z=z|\bX)}\{D-p_z(\mathbf{X})\}+p_z(\mathbf{X}),\\
\psi_{YD,z}=&\frac{\mathbf{1}(Z=z)}{\Pr(Z=z|\bX)}\{YD-m_{z1}(\mathbf{X})p_z(\mathbf{X})\}+m_{z1}(\mathbf{X})p_z(\mathbf{X}).
\end{align*}
Theorem \ref{thm:eif_saces} below characterizes the nonparametric EIF. 
\vspace{-0.1in}
\begin{theorem}[\emph{Efficient influence function}]\label{thm:eif_saces}
The nonparametric EIF for $\mu_{11}$ is $\varphi_{11}(\mathcal{O}|\theta)=E\{e_{11}(\bX)\}^{-1}\left\{\xi^1_{11}(\mathcal{O})-\xi^0_{11}(\mathcal{O})\right\}$, where $\xi_{11}^z(\mathcal{O})=\omega_{11}^z(\mathcal{O})-\mu_{11}^z\tau_{11}(\mathcal{O})$ and
 \begin{align*}
      &\tau_{11}(\mathcal{O})=e_{11}(\bX)+\frac{\theta(\bX)-1}{\sqrt{\delta(\bX)}}
      \sum_{z=0}^1 \left(\psi_{D,z}-p_z(\bX)\right)\left\{\frac{\theta(\bX) p_{1-z}(\bX)}{\theta(\bX)-1}-e_{11}(\bX)\right\},\\
 &\omega^z_{11}(\mathcal{O})=\frac{e_{11}(\bX)}{p_z(\bX)}\left\{\psi_{YD,z}-m_{z1}(\bX)\psi_{D,z}\right\}+\tau_{11}(\mathcal{O})m_{z1}(\bX),~~~~~z\in\{0,1\}.
 \end{align*}
\end{theorem}
For ease of elaboration below, we temporarily assume that the sensitivity parameter $\theta(\bX)=\theta$ does not depend on covariates. We can then treat $\varphi_{11}(\mathcal{O}|\theta)$ as a map of $\theta$ from the extended non-negative real line $[0,\infty]$ to the space of EIFs. By the L'Hôpital's rule, $\lim_{\theta\to\infty}\varphi_{11}(\mathcal{O}|\theta)$ and $\lim_{\theta\to1}\varphi_{11}(\mathcal{O}|\theta)$ (limit is evaluated by treating $\varphi_{11}(\mathcal{O}|\theta)$ as a real number given realizations of $\mathcal{O}$) exist and we denote them as $\varphi_{11}(\mathcal{O}|\infty)$ and $\varphi_{11}(\mathcal{O}|1)$, respectively. On closer inspection, $\varphi_{11}(\mathcal{O}|\infty)$ shares the same expression as $\phi_{11}$ in Theorem 2 of \cite{JiangJRSSB2022}, provided that $p_1(\bX) > p_0(\bX)$ with probability one. Second, under counterfactual intermediate independence, the EIF for $\mu_{11}$ simplifies to
\begin{align*}
\varphi_{11}(\mathcal{O}|1)=&E\{p_0(\bX)p_1(\bX)\}^{-1}\left\{(\psi_{D,0}-p_0(\bX))p_1(\bX)(m_{11}(\bX)-\mu_{11}^1)+ p_0(\bX)(\psi_{YD,1}-\mu_{11}^1\psi_{D,1})\right.\\
&\left.-(\psi_{D,1}-p_1(\bX))p_0(\bX)(m_{01}(\bX)-\mu_{11}^0)-p_1(\bX)(\psi_{YD,0}-\mu_{11}^0\psi_{D,0})\right\}.
\end{align*}
Then the variance of the EIF $\varphi_{11}(\mathcal{O}|1)$ represents the semiparametric variance lower bound for estimating SACE under counterfactual intermediate independence, and the estimator we construct below under this particular case can lead to an improved estimator over the simple weighting estimator proposed by \citet{hayden2005estimator}. 

{Denote $\Gamma(\bX)\equiv[\pi(\bX),p_0(\bX),p_1(\bX),m_{01}(\bX),m_{11}(\bX)]$ as the collection of  nuisance functions. We estimate $\Gamma(\bX)$, denoted as $\widehat{\Gamma}(\bX)$, using either parametric working models or flexible, data-adaptive machine learning approaches.} That is, we solve the estimating equations of the form $\mathbb{P}_n\{\xi_{11}^1(\mathcal{O};\widehat{\Gamma}(\bX))-\xi_{11}^0(\mathcal{O};\widehat{\Gamma}(\bX))\}=0$, which yields an EIF-motivated estimator for the SACE:
\begin{align*}
\widehat{\mu}^{\text{EIF}}_{11}=\displaystyle\frac{\mP_n\left\{\widehat{\omega}^1_{11}(\mathcal{O})-\widehat{\omega}^0_{11}(\mathcal{O})\right\}}{\mP_n\left\{\widehat{\tau}_{11}(\mathcal{O})\right\}},
\end{align*}
where $\left\{\widehat{\tau}_{11}(\mathcal{O}),\widehat{\omega}_{11}^1(\mathcal{O}),\widehat{\omega}_{11}^0(\mathcal{O})\right\}$ represent the fitted values of $\left\{{\tau}_{11}(\mathcal{O}),{\omega}_{11}^1(\mathcal{O}),{\omega}_{11}^0(\mathcal{O})\right\}.$  {We begin with nuisance estimation based on parametric models, following the discussion in Section \ref{s:sace}. The resulting estimator is denoted as $\widehat{\mu}^{\text{CDR}}_{11}$. We summarize the large-sample properties of $\widehat{\mu}^{\text{CDR}}_{11}$ in Theorem S1 of the Supplementary Material, which states that the estimator is conditionally doubly robust. That is, as long as the principal score model is correctly specified, the estimator is consistent if either the propensity score model or the outcome mean model is also correctly specified.} This is consistent with the intuition that the principal score models play a pivotal role in recovering the joint probability mass from correlated counterfactual intermediate outcomes. {Furthermore, Remark S1 of the Supplementary Material summarizes the large-sample properties of $\widehat{\mu}^{\text{CDR}}_{11}$ in the two limiting cases. That is, $\widehat{\mu}^{\text{CDR}}_{11}$ is quadruply robust under the counterfactual intermediate independence assumption, but triply robust under the monotonicity assumption. Furthermore, in Section S3 of the Supplementary Material, we provide details on using a robust sandwich variance estimator to obtain standard errors and interval estimates.}

Although the proposed conditionally doubly robust estimator $\widehat{\mu}^{\text{CDR}}_{11}$ provides some protection against parametric working model misspecification, it may still be biased when all working models are misspecified, even though the sensitivity parameter $\theta(\bX)$ is correctly specified. By leveraging the properties of the derived EIFs, the debiased machine learning approach allows for the effective use of flexible, data-adaptive machine learners for nuisance function estimation \citep{dml}. Specifically, we consider the cross-fitting scheme to implement the EIF-motivated estimator for SACE \citep{pfanzagl1985contributions,klaassen1987consistent}, which offers two advantages: (i) it requires weaker regularity conditions, and (ii) it allows the use of more commonly employed machine learning estimators.

To implement this debiased machine learning estimator, we randomly split the sample into $K$ independent folds, and define $F_i$ as a categorical random variable taking values in $\{1,\ldots,K\}$ representing the fold membership with $\Pr(F_i=k)=K^{-1},1\leq k\leq K$. In practice, with a sufficient sample size, a reasonable choice of  $K$  should satisfy  $n/K \geq 100$. {To improve the stability of the estimation procedure and reduce the risk of imbalanced folds, one may also use stratified sample splitting within stratum defined by treatment $Z$ and the intermediate outcome $D$.}
Let $\mathcal{V}^k$ be the collection of observed data in $k$-th fold, i.e., $\mathcal{V}^k=\{\mathcal{O}_i:F_i=k\}$, and $n_k=\sum_{i=1}^n \mathbf{1}(F_i=k)$ be the sample size of $\mathcal{V}^k$. 
%Notably, the distribution of $F_i$ does not impact the validity of the methods, as long as $n_k/n= 1/K+o_\mP(1)$ and $K$ is finite, and subsequently, one might consider using a deterministic splitting (e.g., nearly equal sample splitting stratified by $Z$ and $D$) to avoid the imbalanced folds when the total sample size is limited. 
For each fold $k$, we estimate all nuisance functions $\Gamma(\bX)$, using the {training} sample $\mathcal{V}^{-k}\equiv\{\mathcal{O}_i:F_i\neq k\}$ ($\mathcal{V}^k$ and $\mathcal{V}^{-k}$ are disjoint samples) by data-adaptive machine learning approaches with convergence rate $n^{1/4}$ in $L_2(\mathbb{P})$-norm; the resulting estimated nuisance functions are denoted as, $\widehat{\Gamma}^k(\bX)\equiv[\whpi^k(\bX),\whp^k_0(\bX),\whp^k_1(\bX),\whm^k_{01}(\bX),\whm^k_{11}(\bX)]$. For example, $\lVert\whpi^k(\bX)-\pi(\bX)\rVert_2=o_{\mathbb{P}}(n^{-1/4})$ can be achieved by numerous well-studied modern machine learning estimators, such as boosting \citep{luo2025high} and random forests \citep{wager2018estimation}. For each {validation} sample $\mathcal{V}^k$ and the corresponding empirical measure, $\mathbb{P}_{k}(V)\equiv n_k^{-1}\sum_{i=1}^n \mathbf{1}(F_i=k)V_i$, we define $\widehat{\omega}^{z,k}_{11}=\mathbb{P}_k\left\{\omega^z_{11}(\mathcal{O};\widehat{\Gamma}^k(\bX))\right\},z\in\{0,1\}$ and $\widehat{\tau}^k_{11}=\mathbb{P}_k\left\{\tau_{11}(\mathcal{O};\widehat{\Gamma}^k(\bX))\right\}$. Here, $\widehat{\omega}^{1,k}_{11}-\widehat{\omega}^{0,k}_{11}$ and $\widehat{\tau}^k_{11}$ are the one-step estimators of $E\{e_{11}(\bX)(m_{11}(\bX)-m_{01}(\bX))\}$ and $E\{e_{11}(\bX)\}$, respectively, based on $\widehat{\Gamma}^k(\bX)$ and $\mathcal{V}^{k}$. To summarize, we use the validation sample to estimate the numerator and denominator of the SACE separately, based on their respective EIFs, after obtaining the estimated nuisance functions from the training sample. Finally, we compute the weighted averages across validation samples and take their ratio to obtain the {debiased machine learning estimator} for $\mu_{11}$:
\begin{align*}
\widehat{\mu}_{11}^{\text{ML}}=\displaystyle\frac{\sum_{{k}=1}^Kn_k\left(\widehat{\omega}^{1,k}_{11}-\widehat{\omega}^{0,k}_{11}\right)}{\sum_{{k}=1}^Kn_k\widehat{\tau}^k_{11}}.
\end{align*}
%Notably, $\widehat{\mu}_{11}^{\text{ML}}$ may be termed as a semi-machine learning estimator (a mixture of parametric and non-parametric) if $\theta\in(0,1)\cup(1,\infty)$, since the principal score functions rely on a parametric working model. In contrast, $\widehat{\mu}_{11}^{\text{ML}}$ is entirely non-parametric if $\theta=0$, $\theta\to1$, or $\theta\to\infty$. 
Theorem \ref{thm:multiply_ml} below shows that $\widehat{\mu}_{11}^{\text{ML}}$ is $\sqrt{n}$-consistent, asymptotically normal and efficient under mild regularity conditions. 

%% FL to continue from here

\begin{theorem}[\emph{Asymptotic property of debiased machine learning}]\label{thm:multiply_ml}
Suppose \\$\text{inf}_{\bX\in\mathcal{X}}\theta(\bX)>0$ and the following regularity conditions hold for some constants $\epsilon_1\in(0,1/2]$ and $\epsilon_2>0$: (1) $\epsilon_1\leq\left\{{\pi}(\bX),p_0(\bX),p_1(\bX),\widehat{\pi}^k(\bX),\widehat{p}^k_0(\bX),\widehat{p}^k_1(\bX)\right\}\leq 1-\epsilon_1$ for any $1\leq k \leq K$, almost surely; (2) For $z\in\{0,1\}$,
    $\{\underset{{1\leq k\leq K}}{\text{max}}\lVert\widehat{\pi}^k(\bX)-\pi(\bX)\rVert_2,\underset{{1\leq k\leq K}}{\text{max}}\lVert\widehat{p}_z^k(\bX)-p_z(\bX)\rVert_2,\underset{{1\leq k\leq K}}{\text{max}}\lVert\widehat{m}_{z1}^k(\bX)-m_{z1}(\bX)\rVert_2\}=o_{\mP}(n^{-1/4});$
    (3) $E\{(Y-m_{z1}(\bX))^2|\bX,Z=z,D=1\}\leq\epsilon_2,z\in\{0,1\}$, almost surely; (4) For $z,z^\prime\in\{0,1\}$,    $\underset{{1\leq k\leq K}}{\text{max}}\lVert(\whpi^k(\bX)-\pi(\bX))m_{z^\prime1}(\bX)\rVert_2=o_\mP(1)$ and $\underset{{1\leq k\leq K}}{\text{max}}\lVert(\whp_z^k(\bX)-p_z(\bX))m_{z^\prime1}(\bX)\rVert_2=o_{\mP}(n^{-1/4})$; (5) $\underset{\bX\in\mathcal{X}^\ast}{\text{inf}}|\theta(\bX)-1|>0$, where $\mathcal{X}^\ast=\{\bX\in\mathcal{X}:\theta(\bX)\neq1\}$. 
Then, $\widehat{\mu}_{11}^\text{ML}-\mu_{11}=\mathbb{P}_n\{\varphi_{11}(\mathcal{O}|\theta)\}+o_{\mathbb{P}}(n^{-1/2})$, which implies that $\widehat{\mu}_{11}^\text{ML}$ is $\sqrt{n}$-consistent and asymptotically linear. Additionally, $\widehat{\mu}_{11}^\text{ML}$ is asymptotic normal and semiparametric efficient with 
$\sqrt{n}(\widehat{\mu}_{11}^{\text{ML}}-\mu_{11})\xrightarrow[]{D}\mathcal{N}(0,E\{\varphi_{11}(\mathcal{O}|\theta)^2\})$, provided that the efficiency lower bound exists such that $E\{\varphi_{11}(\mathcal{O}|\theta)^2\}<\infty$.
\end{theorem}

{Theorem \ref{thm:multiply_ml} excludes the degenerate case $\Pr(\theta(\bX)=0)>0$, which arises if always-takers or never-takers are absent for some $\bX\in\mathcal{X}$ (see Supplementary Material S1 for a further discussion). Extending Theorem \ref{thm:multiply_ml} to handle this case requires an additional regularity condition to ensure $\delta(\bX)$ and $\widehat{\delta}^k(\bX)\equiv\delta(\bX;\whp^k_0(\bX),\whp^k_1(\bX))$ are bounded away from zero (Supplementary Material Regularity condition S1).}
Regularity condition (1) requires the propensity score and principal score functions to be uniformly bounded.
%(the boundedness of $p_z(\bX)$ can be dropped under $\theta=1$). 
Regularity conditions (2) and (4) specify the quality requirements for the nuisance estimators. Regularity condition (4) can be replaced by a stronger (though unrealistic if the support of covariates is unbounded) assumption that $m_{z1}(\bX)$ are uniformly bounded. Regularity condition (5) states that $\theta(\bX)$ may attain a value of $1$ at some covariate values but cannot be arbitrarily close to $1$ over the space of all remaining covariate values. Finally, regularity condition (3) specifies that $m_{z1}(\bX)$ must have a finite variance. Moreover, Theorem \ref{thm:multiply_ml} of the main manuscript and Supplementary Material Theorem S1 imply that $\widehat{\mu}_{11}^\text{ML}$ and $\widehat{\mu}_{11}^\text{CDR}$ are asymptotically equivalent under correctly specified models. 
Finally, we construct a consistent cross-fitted variance estimator for inference: 
\begin{align*}
    \widehat{\mathbb{V}}(\widehat{\mu}_{11}^\text{ML})\equiv \frac{1}{n}\displaystyle \sum_{k=1}^K\frac{n_k}{\left[\widehat{\tau}^k_{11}\right]^2}\mathbb{P}_{k}\left\{\left[\xi^1_{11}(\mathcal{O};\widehat{\Gamma}_k(\bX),\widehat{\mu}^{1,k}_{11})-\xi^0_{11}(\mathcal{O};\widehat{\Gamma}_k(\bX),\widehat{\mu}^{0,k}_{11})\right]^2\right\},
\end{align*}
where $\widehat{\mu}^{z,k}_{11}=\widehat{\omega}_{11}^{z,k}/\widehat{\tau}_{11}^k$. Here, $\widehat{\mathbb{V}}(\widehat{\mu}_{11}^\text{ML})$ is a weighted average of the $K$-empirical versions of $E\{\varphi_{11}(\mathcal{O}|\theta)^2\}$ using $\Gamma^k(\bX)$ and $\widehat{\mu}^{z,k}_{11}$ and its consistency is shown in Section S9 of the Supplementary Material. Thus, the asymptotic variance for $\widehat{\mu}_{11}^\text{ML}$ is given by $n^{-1}\widehat{\mathbb{V}}(\widehat{\mu}_{11}^\text{ML})$, based on which one can construct a normality-based confidence interval.

\section{Simulation study}\label{s:simulation}
\subsection{Simulation design}
\label{s:sim;ss:design}
We conduct simulations to study the performance characteristics of the proposed methods. Our central observations are (i) incorrectly assuming monotonicity when the true data-generating process exhibits non-monotonicity leads to biased inference; (ii) on the other hand, incorrectly assuming non-monotonicity when monotonicity holds in the true data-generating process often exhibits milder bias and undercoverage. Through the simulations, we also investigate the finite-sample performance of the parametric conditionally doubly robust estimator and the debiased machine learning estimator, with empirical verifications of Theorem \ref{thm:multiply_ml} and Supplementary Material Theorem S1. We assume $\theta(\bX)=\theta_{\text{true}}$ is a constant for simplicity and generate $n=500$ units under the following four scenarios: $\theta_{\text{true}}\in\{0.5,2\}$ (representing negative or positive association), $\theta_{\text{true}}=1$ (counterfactual intermediate independence) and $\theta_{\text{true}}=\infty$ constrained by $p_1(\bX)>p_0(\bX)$ (monotonicity holds, and hence the estimators in \citet{JiangJRSSB2022} are expected to be valid). The data-generating process for the full data vector $(Y(0),Y(1),D(0),D(1),Z,\bX)^\top$ is given in the Section S12.1 of the Supplementary Material. We approximate the true PCEs by Monte Carlo methods with a simulated super-population of $10^7$ units. By construction, the true principal score model and the true propensity score model follow logistic regressions, while the true outcome model follows a linear regression. 

We compute four sets of estimators for each data-generating scenario, assuming $\theta \in \{0.5, 1, 2, \infty\}$ that represent both correct and incorrect odds ratio specifications. When assuming $\theta=\infty$, we directly employ the estimator of \citet{JiangJRSSB2022}. {When assuming $\theta=1$, we follow \citet{hayden2005estimator} and the direct specification in Remark \ref{remark:boudanry}. We also obtained results assuming $\theta=0.999999$, which were identical (up to $7$ decimal places) to those under the direct specification, and thus are omitted for brevity.} Specifically, to confirm the observation (i), we compare estimators fitted with $\theta \in \{0.5, 1, 2\}$ against $\theta = \infty$ when $\theta_{\text{true}} \in \{0.5, 1, 2\}$. For (ii), we compare estimators fitted with $\theta = \infty$ against $\theta \in \{0.5, 1, 2\}$ when $\theta_{\text{true}} = \infty$. 

Furthermore, in our simulations, we assume that an analyst may misspecify the nuisance models by using incorrect design matrices for model fitting, akin to the simulation design in \citet{Kang2007STS}. For each set of estimators, we consider five different combinations of working design matrices for the nuisance models: (a) estimating all nuisance models on the original covariates $\mathbf{X}$ (correct parametric model specification with the original design matrix); (b) estimating only the outcome mean model on the transformed covariates $\widetilde{\mathbf{X}}$ (parametric model misspecification with the transformed design matrix); (c) estimating only the propensity score model on $\widetilde{\mathbf{X}}$; (d) estimating only the principal score models on $\widetilde{\mathbf{X}}$; and (e) estimating all nuisance models on $\widetilde{\mathbf{X}}$. Note that the transformation between $\mathbf{X}$ and $\widetilde{\mathbf{X}}$ is smooth, non-linear and one-to-one, as detailed in Section S12 of the Supplementary Material. For the debiased machine learning estimators, we estimate the nuisance functions using the Super Learner \citep{superlearner2007}, which employs an ensemble of algorithms including neural networks \citep{neural_networks2021}, regression trees, and generalized linear models. We set $K=5$ throughout to calculate the debiased machine learning estimator under cross-fitting. We consider the following performance metrics: bias, %Monte Carlo standard deviation (MCSD), 
average empirical standard error (AESE), and empirical coverage probability (CP) using AESE, based on $1000$ simulations. 
%Example \verb"R" code for implementing the proposed methods is available at \url{https://github.com/deckardt98/PCEw-oMonotonicity}. 

{Finally, we conduct an additional set of simulations when the true odds ratio varies with covariates, given by $\theta_{\text{true}}(\bX)=|\bm{1}^\top\bX|+0.1$. In practice, specifying the exact covariate-dependent functional form of $\theta_{\text{true}}(\bX)$ may be challenging. A natural simplification is thus to approximate it by a constant, for example, its expected value $E\{\theta_{\text{true}}(\bX)\}$. To assess the implications of such an approximation, we evaluate our estimators when fitted with $\theta= E\{\theta_{\text{true}}(\bX)\}$, as well as with other constant values that deviate from this expected value. Specifically, assuming the analyst has access to the correct design matrix $\bX$, we also compute the proposed estimators using the true specification, $\theta(\bX)=\theta_{\text{true}}(\bX)$, to serve as a gold-standard. We then compare this gold-standard with estimators that assume $\theta(\bX)$ is misspecified by using one of five constant values in $\{E\{\theta_{\text{true}}(\bX)\},0.5,1,2,\infty\}$ to evaluate the consequence of sensitivity parameter misspecification.}

\subsection{Simulation results}
\label{s:sim;ss:result}
\begin{figure}[ht!]
    \centering
    \includegraphics[scale=0.54]{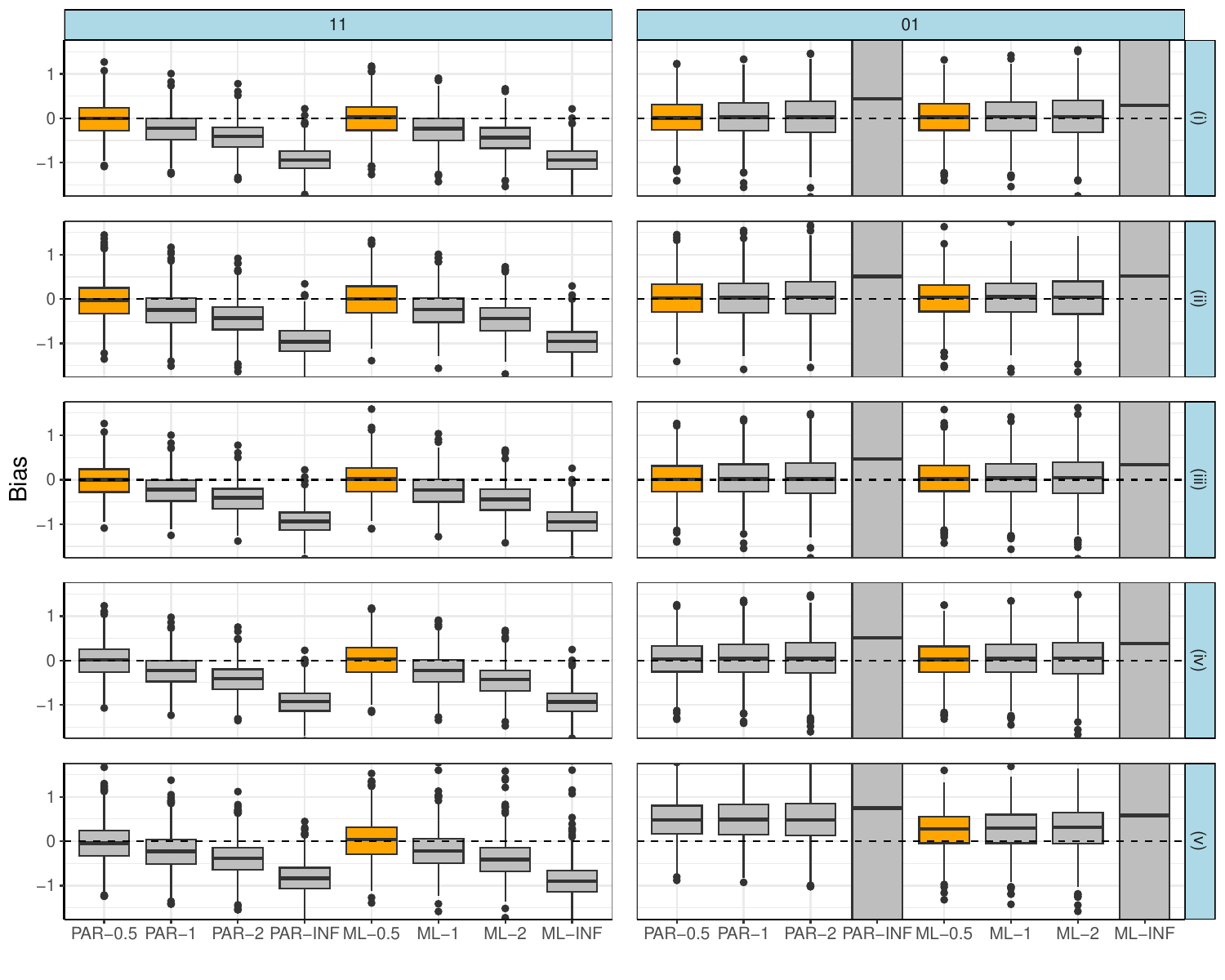}
    \caption{\scriptsize Simulation results for the bias of the average treatment effects among always-takers stratum (`11') and compliers stratum (`01') when the true odds ratio is $\theta_{\text{true}}=0.5$. The design matrices (a)-(e) are described in Section \ref{s:sim;ss:design}. The labels `PAR’ and `DML' represent the parametric conditionally doubly robust estimator and the debiased machine learning estimator, respectively. For instance, `PAR-0.5’ and `PAR-INF’ indicate $\widehat{\mu}_{d_0d_1}^\text{CDR}$ based on assuming $\theta=0.5$ and monotonicity, respectively. The estimators that are expected to be valid are highlighted in orange. }
    \label{fig:sim_theta_0.5_11+01}
\end{figure}

\begin{figure}[ht!]
    \centering
    \includegraphics[scale=0.54]{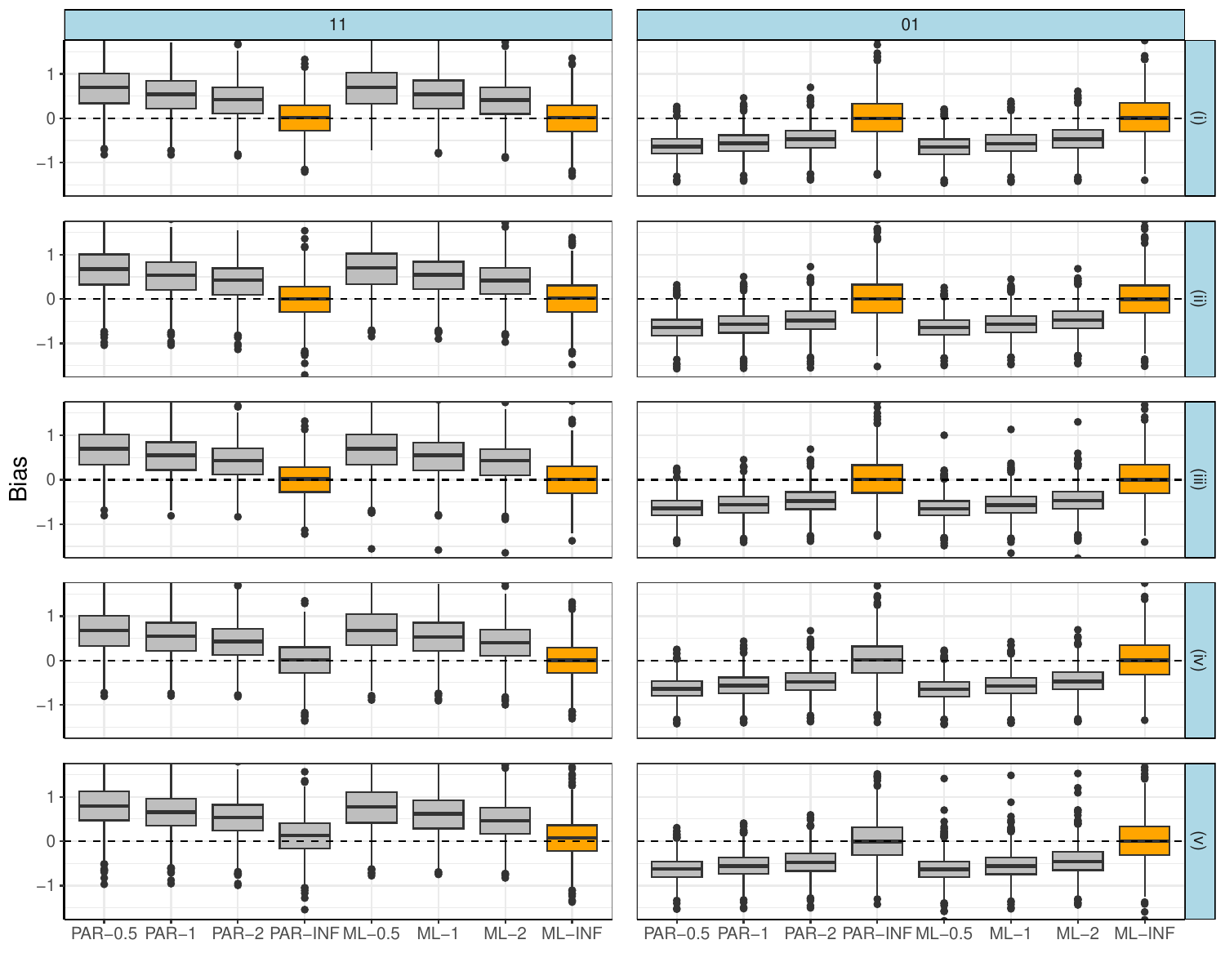}
    \caption{\scriptsize Simulation results for the bias of the average treatment effects among always-takers stratum (`11') and compliers stratum (`01') when monotonicity holds. The design matrices (a)-(e) are described in Section \ref{s:sim;ss:design}. The labels `PAR’ and `DML' represent the parametric conditionally doubly robust estimator and the debiased machine learning estimator, respectively. For instance, `PAR-0.5’ and `PAR-INF’ indicate $\widehat{\mu}_{d_0d_1}^\text{CDR}$ assuming $\theta=0.5$ and monotonicity, respectively. The estimators that are expected to be valid are highlighted in orange. }
    \label{fig:sim_theta_INF_11+01}
\end{figure}

An overview of the key observations and takeaways of the simulation results is summarized in the Supplementary Material Table S7, with elaborations below. To demonstrate observation (i), we start with the simulation results where the true $\theta_{\text{true}}=0.5$, summarized in Figure \ref{fig:sim_theta_0.5_11+01} of the main manuscript, Supplementary Material Figure S1 (box plots of point estimates) as well as Table \ref{tab:sim_CP_0.5} of the main manuscript (empirical coverage probabilities). First, incorrectly assuming monotonicity ($\theta=\infty$) results in the most severe bias of point estimator and undercoverage of interval estimator, compared to the cases with misspecification assuming a finite $\theta$. Second, estimators that incorrectly assume monotonicity are highly unstable and quickly become invalid for estimating the CACE (the $01$ stratum). This instability arises because the denominator of the estimators, $\whp_1(\bX)-\whp_0(\bX)$, is unbounded below or sometimes negative (empirically incompatible principal score estimates). Similar patterns are observed when the true $\theta_{\text{true}} \in \{1, 2\}$, as shown in Supplementary Material Figures S2-S5 and Supplementary Material Tables S4-S5. To demonstrate (ii), we refer readers to the simulation results presented in Figure \ref{fig:sim_theta_INF_11+01} of the main manuscript, Supplementary Material Figure S6 and Supplementary Material Table S6, where monotonicity holds in the data-generating process. Although misspecifying $\theta$ as a finite constant may lead to bias and undercoverage, the impact of such misspecification appears to decrease as $\theta$ increases in magnitude. This suggests that misspecification of $\theta$ as a moderate or large constant may not lead to substantial bias when monotonicity holds. 

Finally, we comment on the operating characteristics of the parametric conditionally doubly robust estimator and the debiased machine learning estimator. When the conditional odds ratio parameter is correctly specified, the parametric conditionally doubly robust estimator with partially correct model specifications, paired with sandwich variance estimators, yields valid inference in finite samples, as evidenced by negligible bias and nominal coverage. Even without a theoretical guarantee, this estimator remains nearly valid when only the principal score model is misspecified. However, this result may have been specific to our chosen data-generating process. When the conditional odds ratio parameter is correctly specified, the debiased machine learning estimator with closed-form variance estimators leads to valid inference, even when the design matrices are transformed by a nonlinear invertible smooth function. This is because the regression function of the transformed covariates remains smooth and regular, allowing data-adaptive machine learners for the nuisance to achieve the nonparametric convergence rate of $n^{-1/4}$. Lastly, the parametric conditionally doubly robust estimator, when all working models are correctly specified, is as efficient as the debiased machine learning estimator using the original correct design matrix. These empirical findings support Theorem \ref{thm:multiply_ml} and Supplementary Material Theorem S1 in the finite-sample context.

{Finally, the additional simulation results when $\theta_{\text{true}}(\bX)$ is covariate-dependent are provided in the Supplementary Material, with bias summarized in Supplementary Material Figures S7-S8 and empirical coverage in Supplementary Material Table S8. As expected, all proposed estimators (with correct design matrix specification) are valid when $\theta(\bX)$ is correctly specified. Nevertheless, misspecification using $\theta=E\{\theta_{\text{true}}(\bX)\}$ (or a close constant) results in only mild bias and undercoverage. This empirical evidence supports the conclusion that if $\theta(\bX)$ does not vary too rapidly with $\bX$, one may still obtain approximately valid inference with a constant approximation, which is a practically feasible option to implement our methods.}

\begin{table}[htbp]
\caption{\scriptsize Simulation results for the (\%) empirical coverage probabilities of 95\% confidence intervals for PCEs, under $\theta_{\text{true}}=0.5$. Results are based on two sets of estimators: the parametric conditionally doubly robust estimator (PAR) and the debiased machine learning estimator (DML), with both correct and incorrect assumptions about $\theta$. The model misspecifications are represented by different design matrices (a)-(e). The symbol ``\textbackslash” indicates that the approach assuming monotonicity is inapplicable for estimating the PCE for defiers. Estimators highlighted in \textbf{bold} are theoretically valid according to asymptotic theory.}
\label{tab:sim_CP_0.5}
\centering
\begin{adjustbox}{width=0.8\textwidth}
%\scriptsize
 \begin{tabular}{ccllllllll}
 \toprule
 ~&~&\multicolumn{4}{c}{PAR-$\theta$}&\multicolumn{4}{c}{DML-$\theta$}\\
  \cmidrule(lr){3-6}  \cmidrule(lr){7-10}
 Strata&Model specification&0.5&1&2&$\infty$&0.5&1&2&$\infty$\\\hline
 \multirow{5}{*}{Always-takers}&(a)& \textbf{94.3} & 89.0 & 72.0 & 11.5 & \textbf{95.9} & 89.7 & 71.2 & 11.7 \\
    ~ & (b)                        & \textbf{93.6} & 89.0 & 76.9 & 23.8 & \textbf{95.0} & 88.8 & 74.4 & 17.9 \\ 
    ~ & (c)                        & \textbf{94.1} & 89.0 & 72.3 & 11.3 & \textbf{95.6} & 89.6 & 70.4 & 12.3 \\ 
    ~ & (d)                        & 93.9 & 88.7 & 72.6 & 11.5 & \textbf{95.2} & 89.7 & 71.0 & 12.2 \\ 
    ~ & (e)                        & 93.7 & 89.2 & 79.3 & 38.1 & \textbf{95.4} & 88.4 & 75.4 & 24.9 \\ 
    \midrule
 \multirow{5}{*}{Compliers}&(a)& \textbf{93.9} & 94.3 & 94.7 & 100.0 & \textbf{94.1} & 94.3 & 94.9 & 97.6 \\
 ~ & (b)                       & \textbf{94.5} & 94.5 & 94.6 & 100.0 & \textbf{94.7} & 94.6 & 94.1 & 97.6 \\ 
 ~ & (c)                       & \textbf{94.3} & 94.6 & 94.6 & 100.0 & \textbf{94.2} & 94.7 & 94.9 & 98.3 \\ 
 ~ & (d)                       & 94.2 & 94.2 & 94.6 & 100.0 & \textbf{94.3} & 94.8 & 95.3 & 98.8 \\ 
 ~ & (e)                       & 81.1 & 82.1 & 84.0 & 99.9  & \textbf{89.3} & 90.8 & 92.4 & 97.9 \\ 
 \midrule
 \multirow{5}{*}{Never-takers}&(a)& \textbf{95.0} & 89.2 & 70.5 & 2.0 & \textbf{96.7} & 90.1 & 69.8 & 2.9 \\
        ~ & (b)                   & \textbf{93.4} & 88.1 & 76.1 & 7.0 & \textbf{94.7} & 89.9 & 74.5 & 10.2 \\ 
        ~ & (c)                   & \textbf{95.1} & 88.9 & 69.9 & 2.1 & \textbf{96.3} & 90.1 & 70.0 & 3.0 \\ 
        ~ & (d)                   & 95.2 & 90.8 & 73.0 & 2.1 & \textbf{96.8} & 91.6 & 73.3 & 2.9 \\ 
        ~ & (e)                   & 92.0 & 83.7 & 66.7 & 8.1 & \textbf{95.2} & 87.9 & 69.3 & 11.9 \\ 
        \midrule
 \multirow{5}{*}{Defiers}&(a)& \textbf{94.5} & 93.9 & 93.2 & \textbackslash & \textbf{94.1} & 93.9 & 94.7 & \textbackslash \\ 
        ~ & (b)              & \textbf{95.0} & 94.7 & 94.4 & \textbackslash & \textbf{94.3} & 94.4 & 94.2 & \textbackslash \\ 
        ~ & (c)              & \textbf{94.2} & 93.9 & 93.7 & \textbackslash & \textbf{94.6} & 94.2 & 94.8 & \textbackslash \\ 
        ~ & (d)              & 94.2 & 93.5 & 93.4 & \textbackslash & \textbf{94.6} & 94.4 & 94.4 & \textbackslash \\ 
        ~ & (e)              & 91.0 & 92.0 & 93.4 & \textbackslash & \textbf{93.4} & 94.5 & 94.9 & \textbackslash \\ 
 \bottomrule
    \end{tabular}
\end{adjustbox}
\end{table}

\section{Analysis of the Acute Respiratory Distress Syndrome trial}
\label{s:Data_Example}
We reanalyze the Acute Respiratory Distress Syndrome (ARDS) Network randomized trial \citep{ARDS04} to illustrate our methods. In this trial, 549 patients with acute lung injury and ARDS who were on mechanical ventilation were randomly assigned to either lower or higher positive end-expiratory pressure (PEEP) levels. During the study period, 68 patients in the lower PEEP arm and 76 patients in the higher PEEP arm died. Our focus is to assess the SACE of PEEP levels ($Z=1$ for lower PEEP level; $Z=0$ for higher PEEP level) on long-term and patient-centered outcomes, specifically the minimum number of days to returning home (DTRH) within the 60-day period, which may not be well-defined if death occurred before the calculation of the final outcome. Since neither the high nor low PEEP levels are considered standard of care and their relative impact on survivor remains controversial, monotonicity is considered less likely to hold in this study, which motivates us to apply our methods. To estimate the nuisance models, we take an elaborate approach and include 25 baseline covariates, including demographic factors like age, gender, and ethnicity, as well as additional clinical variables. Throughout, we assume that $\theta$ is a constant and vary the conditional odds ratio within a plausible range of $\text{log}(\theta)\in[-3,3]$ ($\theta$ is approximately between $0.05$ and $20$). For comparison, we also compute the estimates under monotonicity, by assuming that lower PEEP, compared to higher PEEP, leads to no worse survival in a deterministic fashion. 

\begin{figure}[ht!]
    \centering
    \includegraphics[scale=0.5]{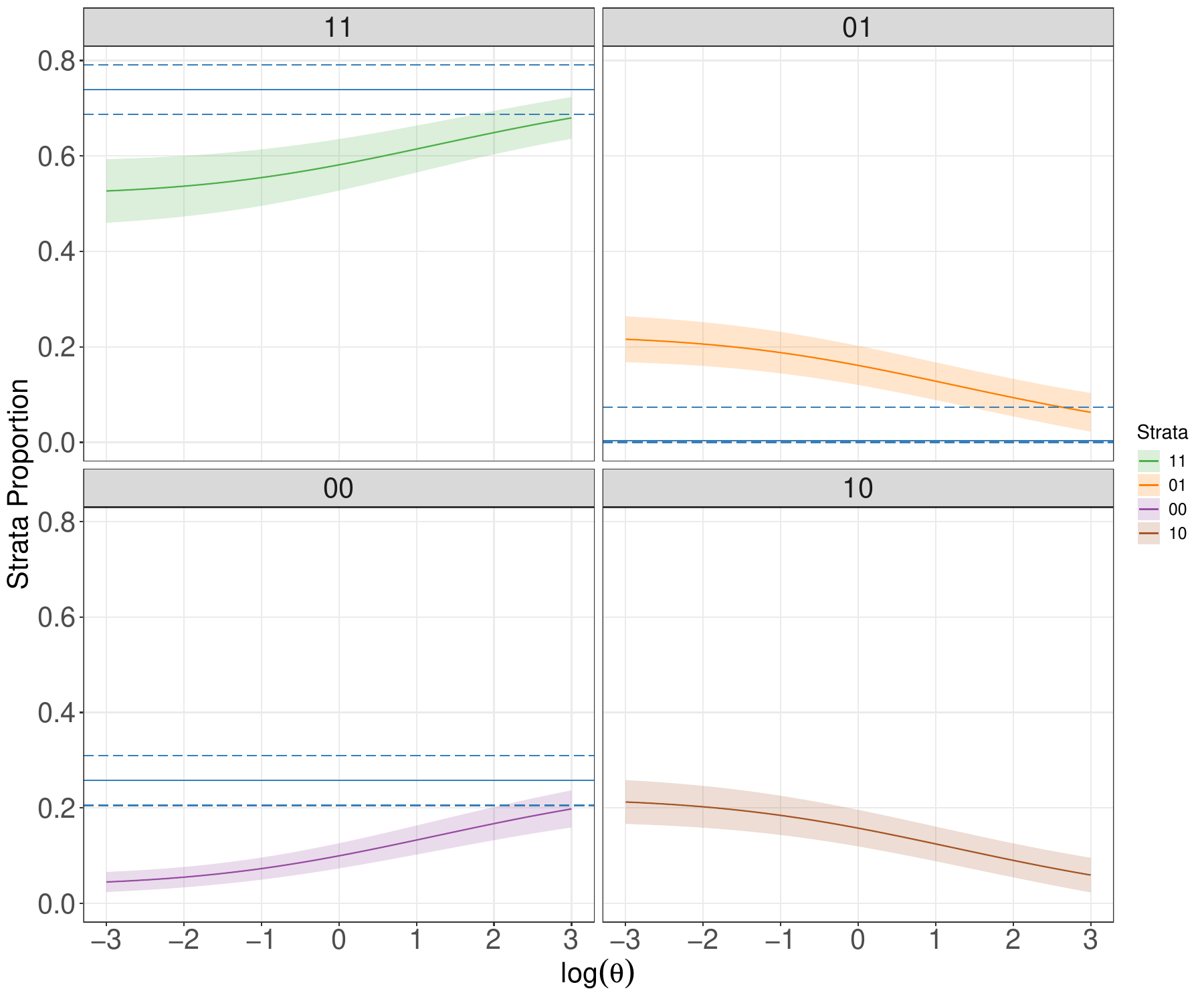}
    \caption{\scriptsize The point estimates and associated 95\% Wald confidence intervals for the strata proportions of the ARDS study are reported, across values of the constant odds ratio sensitivity parameter $\text{log}(\theta) \in [-3,3]$. The strata labels $11,01,00,10$ represent the always-survivors, lower-PEEP-favorable, never-survivors, and higher-PEEP-favorable, respectively. Blue solid and dashed lines indicate the point estimates and confidence intervals under the monotonicity assumption, respectively.}
    \label{fig:ards-ps}
\end{figure}

We first obtain the point estimates and associated 95\% interval estimates for the strata proportions, $e_{d_0d_1}\equiv E\{e_{d_0d_1}(\bX)\}$, using the EIF-based machine learning approach, i.e., $\widehat{e}_{d_0d_1}=n^{-1}\sum_{k=1}^Kn_k\widehat{\tau}^k_{d_0d_1}$, where $\widehat{e}_{d_0d_1}$ is $\sqrt{n}$-consistent, asymptotically normal, and semiparametrically efficient. The results are summarized in Figure \ref{fig:ards-ps}. First, the proportions of the always-survivors ($11$) and never-survivors ($00$) strata increase monotonically as $\theta$ increases, while the trend is reversed for the lower-PEEP-favorable ($01$) and higher-PEEP-favorable ($10$) strata. Second, all strata likely exist for a plausible range of $\theta$, which intuitively aligns with the fact that both treatments are ``active'' (rather than one being standard of care), with no clear superiority on mortality. Thus, crudely assuming monotonicity for the two active treatments tends to overestimate the proportion of always-survivors. Third, the limiting case under monotonicity assumes away the higher-PEEP-favorable strata, but also nearly eliminates the lower-PEEP-favorable stratum ($\lim_{\theta\to\infty} \whe_{01}\approx0.36\%$). Next, we calculate both the parametric conditionally doubly robust estimator and the debiased machine learning estimator for the SACE. Under the parametric approach, we fit logistic regressions for both the propensity score model and the principal score model, and use linear regression for the outcome mean model. For the debiased machine learning approach, we estimate the nuisance functions using the Super Learner \citep{superlearner2007}, given its satisfactory finite-sample performance demonstrated in the simulation study. 

\begin{figure}[ht!]
    \centering
    \includegraphics[scale=0.5]{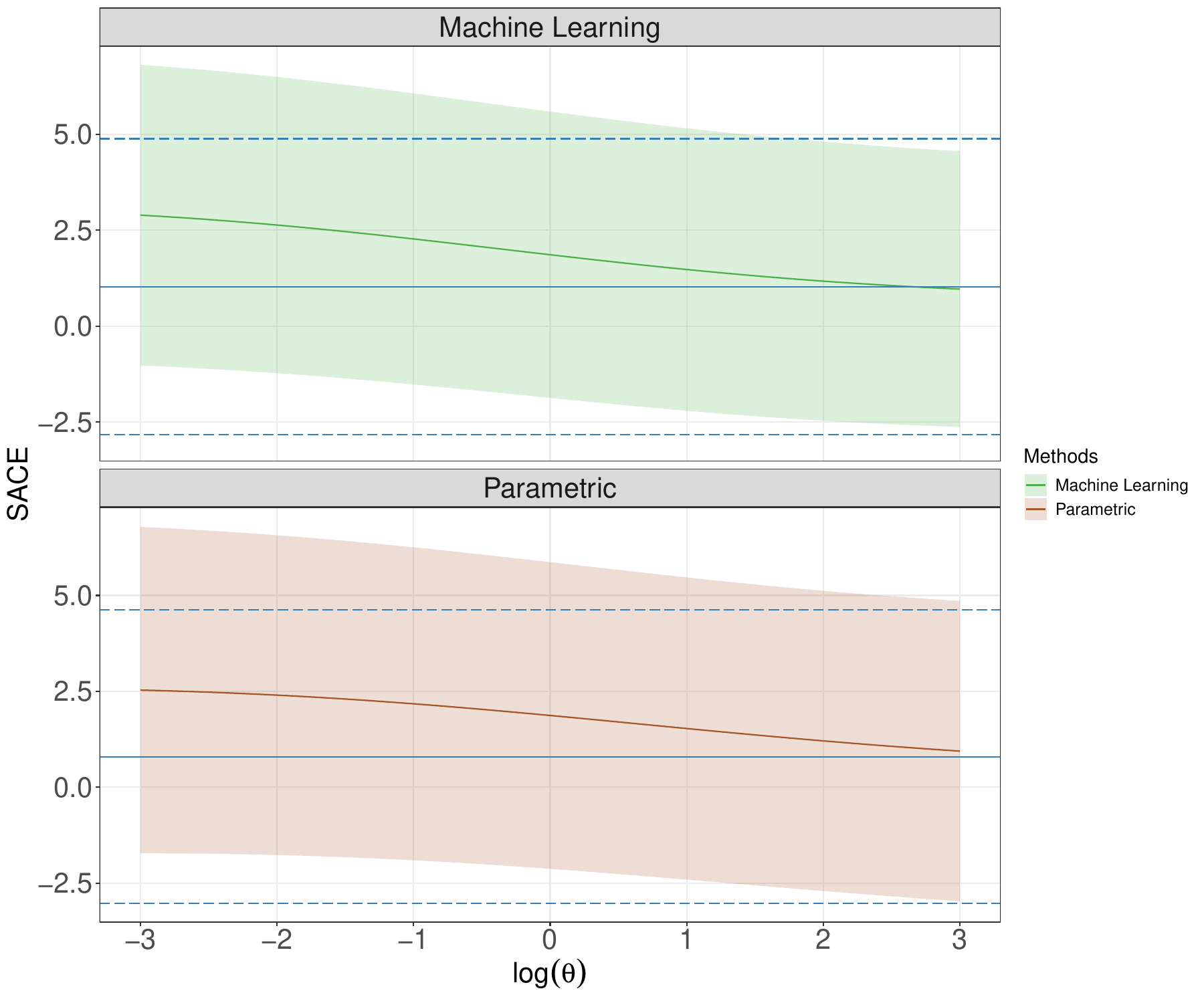}
    \caption{\scriptsize The point estimates and associated 95\% Wald confidence intervals for the parametric conditionally doubly robust estimator and debiased machine learning estimator are reported for the SACE of the ARDS study, across values of the conditional odds ratio $\text{log}(\theta) \in [-3,3]$. Blue solid and dashed lines indicate the point estimates and confidence intervals under the monotonicity assumption, respectively.}
    \label{fig:ards}
\end{figure}

\begin{figure}[htbp]
    \centering
    \includegraphics[scale=0.5]{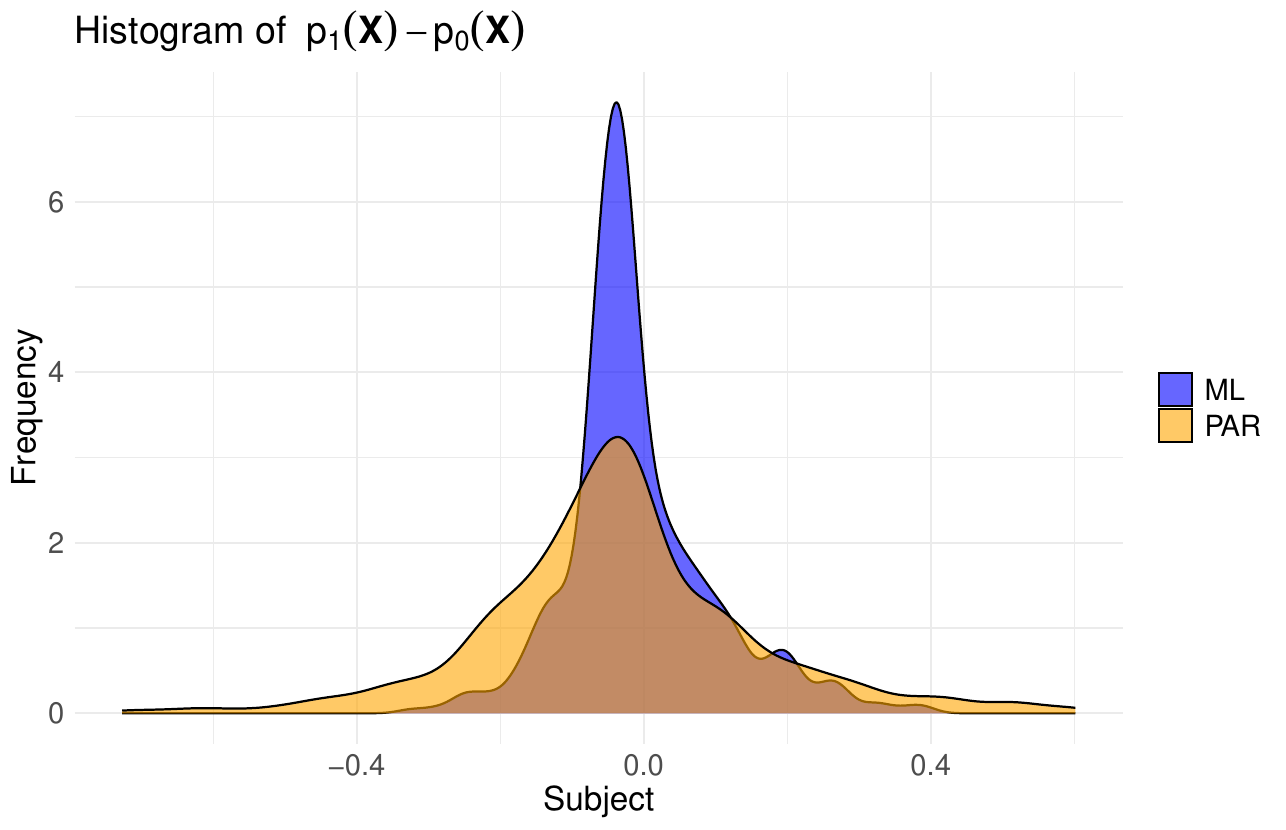}
    \caption{\scriptsize The histogram of of \( \widehat p_1(\bX) - \widehat p_0(\bX) \) across all individuals in the trial application, calculated using Super Learner (‘ML’) and parametric logistic regression (‘PAR’).}
    \label{fig:hist_e10x}
\end{figure}

Figure \ref{fig:ards} shows the point estimates and 95\% confidence intervals for the SACE, using both $\widehat{\mu}_{11}^\text{CDR}$ and $\widehat{\mu}_{11}^\text{ML}$. Several observations emerge. First, the slope for the point and interval estimates with respect to $\theta$ remains negative and shows a clear visual deviation from zero. For example, higher PEEP levels could reduce the DTRH by approximately 2.5 days when $\theta \approx 0.05$, with the effect moderating to about 0.8 days as $\theta \approx 20$. Second, Figure \ref{fig:ards} shows that $\widehat{\mu}_{11}^\text{ML}(\theta)$ intersects with the estimate using the method of \cite{JiangJRSSB2022} at $\theta \approx 16$ and deviates from it as $\theta$ increases. This contradicts the anticipation that $\lim_{\theta\to\infty}\widehat{\mu}_{11}^\text{ML}(\theta)$ should coincide with the estimator in \cite{JiangJRSSB2022} when $p_1(\bX) > p_0(\bX)$ holds with probability one and monotonicity holds. As a further exploration and an empirical check, we plot the histogram of $\whp_1(\bX) - \whp_0(\bX)$ in Figure \ref{fig:hist_e10x} and find that $\whp_1(\bX) - \whp_0(\bX)$ frequently takes negative values, regardless of whether the Super Learner or parametric logistic regression method is used for nuisance estimation. Hence, it is reasonable to question the condition of $p_1(\bX) > p_0(\bX)$ across all units in the trial, and, subsequently, the validity of monotonicity. Third, there is no substantial difference between the parametric and machine learning approaches, suggesting that the parametric working models may reasonably approximate the true data-generating process already in this application. In summary, while statistical significance remains unaffected by potential misspecification of $\theta$, the proposed approach can assess the trend, direction, and magnitude resulting from misspecification of $\theta$.

\section{Discussion}
\label{s:Discussion}

% restate contribution

In this article, we have developed a unified framework for semiparametric principal stratification analyses without enforcing monotonicity, based on a margin-free conditional odds ratio sensitivity parameter; our approach encompasses commonly adopted assumptions like monotonicity and counterfactual intermediate independence as two special limiting cases. We derive non-parametric identification formulas for the principal causal effects and proposed simple weighting and regression estimators, and as an improvement, a conditionally doubly robust estimator based on the EIF. The consistency of the EIF-based estimator requires correct specification of either the propensity score model or the outcome mean model, but not necessarily both, assuming the principal score model is correctly specified; it is also semiparametrically efficient when all working models are correctly specified. Furthermore, we propose a debiased machine learning estimator that is consistent, asymptotically normal, and semiparametrically efficient under mild regularity conditions. We demonstrate that our approaches perform satisfactorily in finite-sample settings through extensive simulations.  We illustrate our methods through a case study based on an acute care trial, where monotonicity is unlikely to hold. Finally, while we focus on relaxing monotonicity, we acknowledge that the validity of the proposed methods may still be compromised if the principal ignorability assumption is violated. To evaluate the impact of deviations from this assumption, we also describe a sensitivity analysis approach for this assumption in Supplementary Material Section S11.

To relax monotonicity in principal stratification analysis, we reiterate the key advantages of the proposed parametrization for sensitivity analysis: (i) the odds ratio is easily interpretable, with its magnitude informed by clinicians’ prior knowledge; (ii) it is margin-free and variation-independent, enhancing practical applicability; and (iii) it unifies two widely used competing assumptions: monotonicity and counterfactual intermediate independence. In our experience, non-monotonicity is not uncommon in trials with two active treatments, for which the proposed methods may be particularly useful. Specifically, our simulations demonstrate that incorrectly assuming monotonicity can pose substantial risks for inference, whereas misspecifying $\theta$ as large constants results in reasonable inference with only mild bias and undercoverage even when monotonicity holds. Additionally, we find that incorrectly assuming monotonicity often lead to unstable estimates for the PCE among compliers, as $\whp_1(\bX) - \whp_0(\bX)$ takes negative values for a certain proportion of units. {In a scenario where the treatment is generally beneficial in a stochastic fashion (unlike the deterministic benefit required by the strict monotonicity assumption), our approach can still provide valid inference by specifying $\theta$ as a large but finite constant. As we discussed in Section \ref{s:simulation}, in practical applications where prior knowledge is limited, specifying $\theta(\bX)$ as a constant remains a useful approximation strategy. In fact, our simulations have shown that a constant approximation near the expected value of the true odds ratio often yields nearly valid inference, especially when $\theta_{\text{true}}(\bX)$ is not dramatically variable with respect to $\bX$. Our application study in Section \ref{s:Data_Example} exemplifies such a constant approximation strategy. At the same time, it is advisable to evaluate the robustness of the results under a range of plausible odds ratios, potentially informed by subject-matter knowledge, discussions with experienced investigators, or evidence from the existing literature.} 

Although we have focused on a binary treatment, it is possible to extend this framework to multiple categorical treatments. The existing literature on principal stratification with multiple categorical treatments with arbitrary levels is relatively sparse \citep[e.g.,][]{tong2024doubly}, due to the challenge of generalizing monotonicity without imposing overly strong assumptions. In theory, the generalization of the proposed framework to multiple categorical treatments may leverage the concept of higher-order odds ratios, as discussed in Chapter 6.2 in \cite{rudas_lectures_2018}. Although the joint probability mass for the latent principal strata does not have a simple explicit formula, it is possible to consider an iterative proportional fitting algorithm for numerical approximation. However, the margin-free and variation-independent properties may be weakened, as discussed in \cite{AOSVariationIndependence2002}. We leave this interesting extension to future work.

\section*{Acknowledgments}

Research in this article was supported by the United States National Institutes of Health (NIH), National Heart, Lung, and Blood Institute (NHLBI, grant numbers R01-HL168202 and 1R01HL178513). All statements in this report, including its findings and conclusions, are solely those of the authors and do not necessarily represent the views of the NIH. The authors also thank the Yale University-Mayo Clinic Center of Excellence in Regulatory Science and Innovation (CERSI) for supporting this study.

\bibliographystyle{chicago}

\bibliography{ref}
\end{document}